\documentclass[12pt]{article}
\usepackage{enumerate}
\usepackage{natbib}
\usepackage{siunitx}
\usepackage[utf8]{inputenc}
\usepackage{amsmath}
\usepackage{amsfonts}
\usepackage{amssymb}
\usepackage{graphicx}
\usepackage{setspace}
\usepackage{caption}
\usepackage{bm}
\usepackage{url}
\usepackage{floatrow}
\usepackage{ulem}
\usepackage{appendix}
\usepackage{csvsimple}
\usepackage{booktabs}
\usepackage{adjustbox}
\usepackage{rotating}
\usepackage{url} 

\newcommand{\blind}{1}

\begin{document}

\def\spacingset#1{\renewcommand{\baselinestretch}%
{#1}\small\normalsize} \spacingset{1}


\if1\blind
{
  \title{\bf A Powerful Modelling Framework for Nowcasting and Forecasting COVID-19 and Other Diseases}
  \author{Oliver Stoner\thanks{
    This work was supported by an EPSRC Doctoral Training Partnership studentship awarded to Alba Halliday.}\hspace{.2cm}, 
    Theo Economou
    and 
    Alba Halliday\\
    Department of Mathematics, University of Exeter, UK}
  \maketitle
} \fi

\if0\blind
{
  \bigskip
  \bigskip
  \bigskip
  \begin{center}
    {\LARGE\bf A Powerful Modelling Framework for Nowcasting and Forecasting COVID-19 and Other Diseases}
\end{center}
  \medskip
} \fi

\bigskip
\begin{abstract}
The COVID-19 pandemic has highlighted delayed reporting as a significant impediment to effective disease surveillance and decision-making. In the absence of timely data, statistical models which account for delays can be adopted to nowcast and forecast cases or deaths. We discuss the four key sources of systematic and random variability in available data for COVID-19 and other diseases, and critically evaluate current state-of-the-art methods with respect to  appropriately separating and capturing this variability. We present a general spatio-temporal hierarchical framework for correcting delayed reporting and demonstrate its application to daily English hospital deaths from COVID-19 and Severe Acute Respiratory Infection cases in Brazil. We compare our approach to competing models with respect to theoretical flexibility and quantitative metrics from a rolling nowcasting experiment imitating a realistic operational scenario. Based on consistent and compelling leads in nowcasting accuracy, bias, and precision, we demonstrate that our approach represents the current best-practice for correcting delayed reporting.
\end{abstract}

\noindent%
\vfill

\newpage
\spacingset{1.5} 

\section{Introduction}\label{sec:intro}
The coronavirus disease or COVID-19 is an infectious disease caused by the severe acute respiratory syndrome coronavirus 2 (SARS-Cov-2) virus. Like many infectious diseases, data on COVID-19 cases and deaths are typically subject to delayed reporting, otherwise known as `notification delay'. This is when available count data are, for a time, an under-representation of the truth, owing to flaws or `lags' in the data collection mechanism. In disease surveillance, delays -- e.g. ones that occur during the transfer of information from local clinics to national surveillance centres -- mean that complete and informative counts of new cases or deaths are not immediately available. Often these delays are substantial, so that it can take several weeks or even months for the available data to reach a total reported count. 
\begin{figure}[h!]
\floatbox[{\capbeside\thisfloatsetup{capbesideposition={right,center},
capbesidewidth=0.5\linewidth}}]{figure}[1\linewidth]
{\caption{Bar plot of reported COVID-19 hospital deaths in the East of England region, for the days leading up to and including day $t$, the 4th of May 2020. The grey bars represent the total (as yet unobserved) number of reported deaths, while the different coloured bars show the number of deaths reported  after each day of delay.} \label{fig:bar_plot}}
{\includegraphics[width=\linewidth]{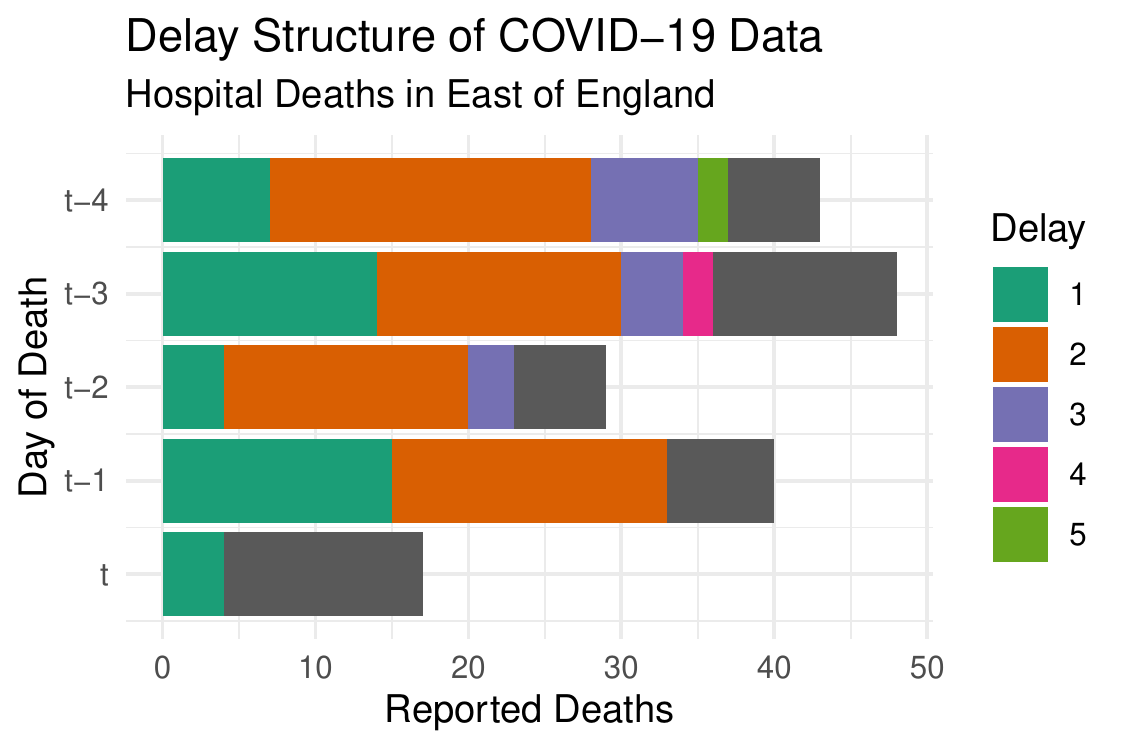}}
\end{figure}

To visualise the data challenge, Figure \ref{fig:bar_plot} shows reported COVID-19 hospital deaths in the East of England in the days leading up to and including day $t$, the 4th of May 2020. The coloured bars show the partially reported data available at the end of day $t$, while the grey bars show the as-of-yet unknown total number of deaths for each date. For day $t$, we have only observed the portion of deaths reported within the latest 24 hour reporting period (here referred to as `within the first delay' and shown in green). For $t-1$, we have data reported within the first delay (green) as well as ones reported within the second delay (orange). The number of available `delayed' counts therefore increases by one for each day we go back into the past.

Significant changes in the delay mechanism (e.g. in the proportion of deaths reported in the first delay) make it challenging to draw conclusions about the total counts in a timely manner. For example, a similar number of deaths were reported within the first delay for days $t$ and $t-2$ (Figure \ref{fig:bar_plot}), while the total number of deaths (grey) was much higher for $t-2$. For a practitioner analysing the data at the end of day $t$, there is no clear way of knowing from the available reported counts that the total for day $t$ is comparatively low -- especially given the increased death count reported so far occurring on day $t-1$ compared to day $t-2$. In disease surveillance, therefore, delayed reporting can make it difficult to confidently detect an outbreak within a time frame during which interventions are most effective. A particular issue during the COVID-19 pandemic is the need to confidently detect any local outbreaks so that interventions like increased social distancing measures can be considered and implemented. Here, failure to act in a timely manner carries the risk of loss of life, while unnecessary interventions can also be costly for the local economy or other aspects of population well-being.

From a statistical perspective, tackling delayed reporting is a prediction problem. We would like to predict (nowcast) the present-day total count (e.g. the number of deaths), as well as forecast future counts, based on any available partial counts and potentially on any historical total counts which have now been fully observed. In this article we propose and evaluate a compelling spatio-temporal hierarchical approach to correcting delayed reporting in COVID-19 data and other disease surveillance applications. The article is structured as follows: in Section \ref{sec:background}, we discuss the need to consider different sources of variability in COVID-19 data suffering from delayed reporting and use this as a principled basis for comparing existing approaches; in Section \ref{sec:method}, we present our general framework for correcting delayed reporting in COVID-19 data, alongside a general discussion of spatial, temporal,  and spatio-temporal structures which may be included in the model; in Section \ref{sec:app}, we apply this framework to counts of hospital deaths from COVID-19 in regions of England and present a rolling prediction experiment to illustrate the model's operational effectiveness in comparison with other approaches. Finally, we conclude with a critical discussion of our approach and avenues for future research in Section \ref{sec:discuss}.

Accompanying the article is a substantial appendix, which is structured as follows: in Appendix A we apply our approach to Severe Acute Respiratory Infection (SARI) data from Brazil, to both illustrate the framework's applicability to general disease surveillance data suffering from delayed reporting and also to compare independent time series models for each region with a joint model sharing the same spatio-temporal structure as the COVID-19 application; in Appendix B we present the mathematical formulation of competing models appearing in section \ref{sec:app}; and finally, in Appendix C we discuss how our framework could be adjusted to take into account under-reporting in the final reported death/case counts.

\section{Background}\label{sec:background}
\begin{figure}[h!]
\caption{Scatter plot of daily hospital deaths in England. Dashed line and points: the number of deaths reported on each day (announced deaths). Different shapes and colours represent the day of the week. Dotted line: the number of actual deaths on each day ($y$). Solid line: smooth trend of the actual deaths.} \label{fig:new_deaths}
\includegraphics[width=\linewidth]{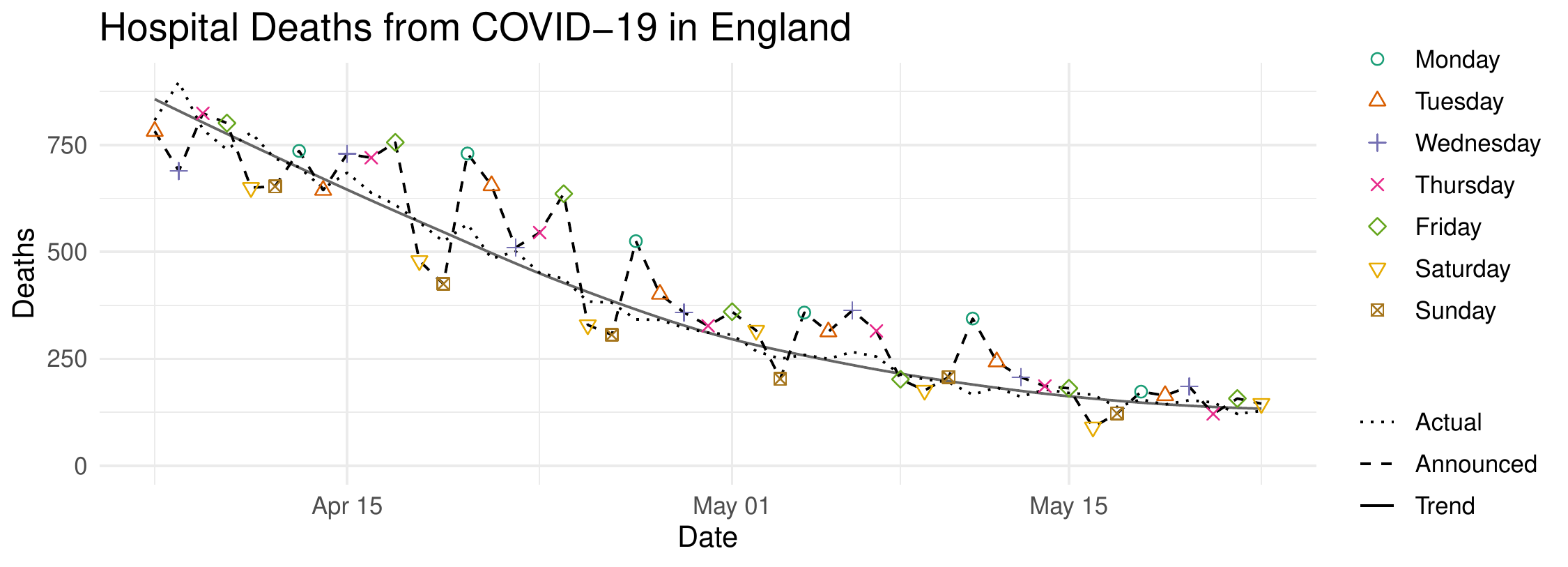}
\end{figure}

We begin by introducing some notation. Let $y$ be the total count, e.g. the number of COVID-19 deaths or cases occurring on a given day, and let $z_{d}$ be the portion of $y$ observed within $d = 1,\dots,D$ delays, so that $\sum_{d=1}^{D} z_{d} = y$. To better understand existing modelling approaches, it is instructive to appreciate the different sources of variability which might be present in data relating to COVID-19 but also other diseases. Figure \ref{fig:new_deaths} shows the total number of COVID-19 in-hospital deaths that were identified on each day in England, which we call the `announced' deaths. These are deaths that were confirmed on that day, but may have occurred days ago -- shown here in order to evidence the variability in the reporting delay. Also plotted are the `actual' deaths ($y$) confirmed to have occurred on that day (which is unknown for a time due to delayed reporting). The period in concern (April-May) is broadly after the first peak in the UK, and there is a clear downward trend in both the actual and announced deaths. This downward trend illustrated by the solid line is what we call the `systematic variability' in $y$, which will vary regionally e.g. due to different population sizes, population densities or time since disease took hold of the region. The day-to-day fluctuation of the actual death count $y$ about the smooth curve, is what we call the `random variability' of $y$.

In addition to the variability in $y$, we must also consider variability in the reporting delay, which can also be decomposed into random and systematic. Notice for instance the clear `weekly cycle' in the announced deaths (Figure \ref{fig:new_deaths}) -- also referred to as the `weekend effect' -- where significantly fewer deaths are announced at weekends. The weekend effect for this data can be explained by lower levels of administrative staffing at many hospitals trusts on Saturday and Sunday. This further highlights the issue in using such data as a raw indicator of the progression of the epidemic, especially when flaws like delayed-reporting aren't always communicated transparently. From a modelling perspective, failure to take into account this kind of systematic variability in the reporting delay (which in conjunction with variability in $y$ makes up the variability in the delayed counts $z_d$) means ignoring crucial information when it comes to nowcasting and forecasting.

In addition to the weekly cycle, we would also expect systematic between-region variability in the reporting delay, e.g. resource inequality between regions; as well as systematic temporal variability, e.g. if reporting efficiency improves over time. In summary, attempts to correct for delayed reporting of COVID-19 should carefully consider the following four sources of variability in the available data:
\begin{description}
\item[(a)] Systematic variability in the total count $y$ (e.g. exponential growth/decay, seasonal patterns, regional variation).
\item[(b)] Random variability in $y$ (e.g. day-to-day variation in death count).
\item[(c)] Systematic variability in the reporting delay (e.g. weekly cycles, improvements in reporting efficiency over time, between-region differences).
\item [(d)] Random variability in the reporting delay (e.g. day-to-day variation).
\end{description}

The available data at any given time comprise historical (fully) reported counts $y$ and partial counts $z_d$ corresponding both to historic $y$ and to more recent unobserved $y$. These are the sources of information to be utilised for nowcasting and forecasting and as explained in the subsequent section, the appropriate handling of their respective variability will result in more optimal predictions of current and future counts $y$.

\subsection{Review of existing approaches}
\cite{DR} present an overview of the well established biostatistical literature on correcting reporting delay. Here we revisit some of that but with a particular focus on utility to COVID-19 applications: \cite{hohle2014bayesian} and \cite{salmon1} both propose approaches which combine a Poisson/Negative-Binomial model to describe $y$ with a Multinomial model for the partial counts $z_d|y$, to describe variability in the delayed reporting. These models are applied to Shiga toxin-producing Escherichia coli (STEC) \citep{hohle2014bayesian} and Salmonella \citep{salmon1} data, respectively. The main strength of these approaches is the intuitive separation of variability (random and systematic) in the total count $y$ (a \& b) from variability in the reporting delay (c \& d). In particular, \cite{hohle2014bayesian} present two separate options: 1) the Multinomial probabilities are realisations from the Generalized-Dirichlet distribution for each time step, and 2) the Multinomial probabilities are modelled with a logistic transformation of potentially informative covariates. The first option offers considerable flexibility to capture different amounts of random variability in the reporting delay, but lacks the capability of capturing systematic variability like the weekly-cycle in reporting performance discussed previously. Meanwhile, the second option allows such systematic variability to be captured, at the expense of model fit and non-optimal predictions in the (very common) situations where delayed counts $z_d|y$ are over-dispersed with respect to the Multinomial \citep{DR}.

Epidemiological applications (including disease surveillance) often have a spatial dimension \citep{spatio_dengue} and this is certainly true for COVID-19, where data are often grouped into geographical units like regions or health authorities. Two existing approaches which deal with spato-temporal data are \cite{theo_dengue} and \cite{thailand}. In both cases, the partial counts $z_d$ are assumed Negative-Binomial in a Bayesian hierarchical framework, where $\mathbb{E}[z_d]=\mu_d$ depends on covariates and random effects intended to capture systematic variability in the total count (a) -- albeit indirectly through $y=\sum_d z_d$ -- and in the reporting delay (c). This approach, applied to spatio-temporal SARI data from Brazil \citep{theo_dengue} and to dengue fever data from Thailand \citep{thailand}, is a generalisation of older chain-ladder approaches (e.g. \cite{mack_1993}) and is quite flexible, as it can potentially incorporate a wide variety of temporal, spatial and spatio-temporal structures. However, the total counts are not explicitly modelled, while the partial counts are assumed independent given covariates and random effects. As such, random variability in the total counts (b) is not necessarily captured well in addition to the added risk of excessive predictive uncertainty when nowcasting and forecasting \citep{DR}. This is in part due to the lack of separation between systematic variability in the total count (a) and the reporting delay (c). A similar approach which partly addresses this separation issue is given by \cite{nobs}, where the mean of $z_d$ is defined as $\mu_d=\beta_d\lambda$. Parameter $\beta_d\in(0,1)$, where $\sum_d\beta_d=1$, is the proportion expected to be reported with delay $d$, while $\lambda=\mathbb{E}[y]$ is effectively the mean of the total count. The proportions $\bm{\beta}=\{\beta_d\}$ are fixed, while $\lambda$ is modelled by random effects at the log-scale. To account for systematic variation (over time) in the reporting delay (c), the model is applied over a sliding temporal window of fixed length. As such, $\beta_d$ is representative of reporting behaviour in more recent data. Although this allows flexibility to capture structured temporal variability in the delay, it may result in over-smoothing of the delay distribution if the window size is too big relative to significant short-term structured variability in reporting performance (like those exhibited by UK COVID-19 data, as illustrated later in Figure \ref{fig:weekly}). 

Finally, \cite{DR} propose and assess a general framework for correcting delayed reporting, which utilises a Negative-Binomial model for $y$ and a Generalized-Dirichlet-Multinomial (GDM) model for $z_d|y$. Covariates and random effects can be included in the parameters of the GDM, to account for systematic variability in the mean and variance of the reporting delay. The benefit of this approach is that all four sources of variability are accounted for separately using flexible distributions, leading to enhanced interpretability of the model design along with improved prediction performance when nowcasting and forecasting \citep{DR}. In the following sections we will detail how a spatio-temporal extension of this framework can be used to correct delayed reporting in COVID-19 (Section \ref{sec:app}) and other disease surveillance data (Appendix A).

The various approaches can be broadly classified in two groups: one where the delayed counts $z_d$ are modelled marginally without explicitly modelling/using historical information on the totals $y$, e.g. \cite{theo_dengue} and \cite{nobs}; and another which models the delay counts jointly but also conditionally on $y$, i.e. $\bm{z}|y$, in conjunction with a separate model for $y$, e.g. \cite{DR} and \cite{hohle2014bayesian}. We argue that the latter approach is better able to explicitly capture (a) and (b) in the model for $y$, as well as (c) and (d) in the model for $\bm{z}|y$, especially if the model is sufficiently flexible to allow for overdispersion relative to the multinomial -- like the GDM. Emphasising that the predictand of interest is the total count $y$, we note that the GDM framework implemented in the Bayesian framework produces the predictive distribution $p\left(y^{(\mbox{\scriptsize{unseen}})}|y^{(\mbox{\scriptsize{obs}})},\bm{z}^{(\mbox{\scriptsize{obs}})}\right)$, thus utilising all available information. The `marginal' approaches predict $y$ indirectly as $\sum_d z_d$, thus needlessly predicting unseen $z_d$'s whose uncertainty will propagate in the predicted $y$ and potentially failing to appropriately capture the random variability of $y$, due to the lack of an explicit model for historical $y$.

In Section \ref{sec:compare} we apply models based on our GDM approach, \cite{theo_dengue} and \cite{nobs} to COVID-19 death data from the UK, and make comparisons based on nowcasting and forecasting performance, as well as interpretability. We compare these three approaches as, in our opinion, they are the main three contenders (in terms of flexibility and operational applicability) for operational COVID-19 delay correction.

\section{Modelling framework}\label{sec:method}
Extending the GDM framework in \cite{DR} to include a spatial dimension $s \in S$ (e.g. districts, regions, countries) results in the following mathematical formulation of the model:
\begin{eqnarray}
y_{t,s}\mid \lambda_{t,s}, \theta_s &\sim & \mbox{Negative-Binomial}(\lambda_{t,s},\theta_s); \qquad\log(\lambda_{t,s}) =  f(t,s);\label{eq:nb}\\
\bm{z}_{t,s} & \sim & \mbox{GDM}(\bm{\nu}_{t,s},\bm{\phi}_{t,s},y_{t,s}).
\end{eqnarray}
Systematic spatio-temporal variability in the total counts $y_{t,s}$ is captured by the general function $f(t,s)$, which may include an offset (e.g. population), covariates or random effects. Variability in the delay mechanism is modelled by the Generalized-Dirichlet-Multinomial (GDM) distribution, a Multinomial mixture whose vector of probabilities has a Generalized-Dirichlet distribution \citep{generalizeddirichlet}. The use of the GDM for modelling the partial counts, instead of the more conventional Multinomial, affords a great deal of extra flexibility in accounting for over-dispersion in the random variability of the reporting delay (d) -- which improves nowcasting efforts -- and in capturing unusual covariance structures in the partial counts \citep{DR}. Here we choose to parametrize the GDM in terms of $\bm{\nu}_{t,s}$ and $\bm{\phi}_{t,s}$. These are respectively the mean and dispersion parameters of the Beta-Binomial conditional model for each partial count:
\begin{eqnarray}
z_{t,s,d}\mid \bm{z}_{t,s,-d},y_{t,s} &\sim & \mbox{Beta-Binomial}(\nu_{t,s,d},\phi_{t,s,d},n_{t,s,d}=y_{t,s}-\sum_{j< d} z_{t,s,j}). \label{dr:eq:beta-binomial}
\end{eqnarray}
Parameter $\nu_{t,s,d}$ (the relative mean) is therefore the proportion of the yet-to-be unreported part of $y_{t,s}$ which is expected to be reported at delay $d$. In \cite{DR}, two options were suggested for modelling the relative means $\nu_{t,s,d}$. In the first (named the Hazard variant) they are modelled directly with a logit link, so that $\log\left(\nu_{t,s,d}/(1-\nu_{t,s,d})\right) = g(t,s,d)$, for some general function $g(t,s,d)$. In the second (the Survivor variant) a model is first constructed for $S_{t,s,d}$, the expected cumulative proportion reported after delay $d$:
\begin{eqnarray}
\mbox{probit}(S_{t,s,d})=g(t,s,d).
\end{eqnarray}
The relative means are then easily derived as $\nu_{t,s,d} = (S_{t,s,d}-S_{t,s,d-1})/(1-S_{t,s,d-1})$. \cite{DR} argue that it is more intuitive to consider models for the cumulative proportion of $y_{t,s}$ reported by delay $d$, than to consider models for the expected proportion of $y_{t,s}$ reported at delay $d$ out of those not already reported by delay $d-1$, and so advocate adoption of the Survivor variant over the Hazard variant. The operational characteristics and performance of both options have been studied extensively in \cite{DR}. In the next subsection we discuss how general functions $f(t,s)$ and $g(t,s,d)$ may be appropriately specified in a spatio-temporal context, specifically for the Survivor variant of the GDM framework.

\subsection{Spatio-temporal variation}\label{sec:spatial}
In epidemiological applications like surveillance of COVID-19, we often seek to take into account both structured and unstructured spatial variability in what we consider here to be the total counts (i.e. COVID-19 cases or deaths). For example, COVID-19 cases and deaths may be correlated across neighbouring regions due to population movement. We may also need to consider how spatial and temporal variability might interact. For example, a COVID-19 outbreak might be concentrated in a cluster of regions initially, but spread to more regions as time progresses. With data plagued by reporting delays, we should also account for spatio-temporal variability in the delay structure. For example, data from some regions may be available more quickly than data from others, and structured temporal variability in the delay mechanism may vary with space, e.g. due to changes over time in surveillance resourcing within individual regions.

To motivate these efforts, it is worth considering the reasons for which joint modelling of the data across regions is needed, as opposed to applying a time series model to each region separately. One possible situation is a disease surveillance system where the sum of cases/deaths over a number of regions is important, e.g. for planning resource allocation (e.g. tests, ventilators) on a larger geographical scale. Even if the joint (i.e. spatio-temporal) and independent (i.e. individual time series) models perform equally well at capturing the variance of the total counts in each region, the risk of not capturing similarities across multiple regions (e.g. in their temporal trends) is that the variance of any sum $V(S')=\mbox{Var}\left[\sum_{s\in S'} y_{t,s} \right]=\sum_{i\in S'}\sum_{j \in S'} \mbox{Cov}[y_{t,i},y_{t,j}]$ for some $S'\subseteq S$, may not be captured well -- as we illustrate in Appendix A. A well-specified joint model is potentially able to explicitly quantify the covariance of $y_{t,s}$ across regions (at least at the mean level which may indeed be sufficient), so that $V(S')$ is captured better. Meanwhile, for applications with a high spatial resolution (e.g. local authorities), incorporating spatio-temporal structures may enable a better understanding of how the disease is spreading, to allow resources to be allocated to areas which are likely to be affected in the near future. Finally, cases where missing information arises from sources other than reporting delays, e.g. data loss or national holidays, jointly modelling the regions is important so that regions with less data can potentially borrow information from the others.

The selection of appropriate ways of capturing spatio-temporal variability is well-established in the field of epidemiology. In disease data, space is often defined by areal units (e.g. regions, counties etc.) although sometimes data occur as points in space (e.g. counts of cases at individual clinics or hospitals). In either case, a sensible starting point in defining the model for the mean incidence rate, $f(t,s)$, is to consider separable functions of the form:
\begin{equation}
f(t,s) = f_1 (t) + f_2 (s) + f_3 (t,s).
\end{equation}
Here $f_1(t)$ allows for any common temporal or seasonal variation across the regions; $f_2 (s)$ (which may include offsets such as population) allows for any overall differences in the mean of $y_{t,s}$ between regions; and $f_3 (t,s)$ allows for spatio-temporal interactions.

For the purposes of this article we need not restrict ourselves to any specific models for capturing spatio-temporal variability, among the many that are available (see for instance \cite{banerjee}), particularly as this choice is often application dependent. Instead we will briefly give some illustrative examples and then discuss in more detail an approach based on nested spline structures, in our applications to COVID-19 and SARI data.

Formulating the model for the expected cumulative proportion through $g(t,s,d)$ is slightly more complex due to the extra dimension, but can be decomposed in a similar manner, noting that in practice, $d$ is discrete and bounded from above. Here we consider separable functions of the form:
\begin{equation}
g(t,s,d) = g_1 (t) + g_2 (s) + g_3(d) + g_4 (t,s) + g_5 (t,d) + g_6(s,d) + g_7(t,s,d).
\end{equation}
This allows for spatio-temporal variability in the delay mechanism, e.g. by including tensor product smoothing terms of time and delay which vary with space \citep{tensor}.

It is important to appreciate the need for structures involving interactions between space, time and delay, and to consider whether these make sense within the context of the application and the data. For instance, the delay mechanism may not, in some applications, vary across space $s$ where this is defined at a small spatial scale (e.g. individual clinics or hospitals), but it may well vary across $s$ at a larger scale (e.g. counties/regions). In the application to COVID-19 mortality data in Section \ref{sec:app}, we opt for choices of $f(\cdot)$ and $g(\cdot)$ that are flexible yet practically feasible and appropriate to the application.

\section{Application to COVID-19 deaths}\label{sec:app}
The National Health Service for England (NHS England) publishes daily count data of deaths occurring in hospitals in England of patients who had either tested positive for COVID-19 or where COVID-19 was mentioned on their death certificate. Each daily file presents deaths reported in the time period from 4pm 2 days prior to publication until 4pm on the day before publication, grouped in time by date of death and in space by region (e.g. South West England) but also higher resolutions like NHS Trusts. Assembling published files for each day over a period of time then provides information on the reporting delay by constructing, for each date of death and geographic region, the number of deaths reported on each day. For more straight-forward statistical analysis and compatibility with existing modelling approaches, the number of deaths reported on each day can be organised by the number of days of delay relative to the date of death, taking the first delay ($d=1$) to be deaths which occurred and were reported within the latest reporting period. 

As with other approaches (e.g. \cite{nobs}), the total counts must be assumed fully reported after a specified cut-off, which we denote $D_{max}$. Resulting predictions of $y$ can then be interpreted as the number of cases/deaths which will be reported after $D_{max}$ days from the actual day of death. If only a low proportion (e.g. $<50\%$ of $y$ tends to be reported after $D_{max}$, then nowcasts and forecasts will not offer a complete picture of ongoing or upcoming outbreaks to decision-makers. If $D_{max}$ is needlessly high, then more data on totals $y$ will be unknown and thus require sampling during model fitting, increasing the complexity of the model and potentially making the model impractical for real-time (e.g. daily) prediction. Ideally, $D_{max}$ is chosen to be sufficiently high that on average most of $y$ (e.g. 90\%) are reported. The choice of $D_{max}$ is therefore very application-dependent but not daunting, because in many applications most of $y$ is reported in the first few delays (i.e. $d<10$), with less and less reported afterwards. For this dataset, over 90\% of all deaths reported within 28 days -- in the time period from 2nd of April 2020 to 24th of July 2020 -- were reported within 7 days and over 95\% were reported within 14 days. Here we opt for $D_{max}=14$ days. If no value of $D_{max}$ is specified, then all total counts $y$ are unknown and the model is non-identifiable without additional information (e.g. informative prior distributions), similar to the case of correcting under-reporting \citep{CUR}.

\subsection{Nested spline model}\label{sec:nested}
\cite{DR} present a model for a time series of dengue fever data in Rio de Janeiro, Brazil, where the incidence of the total recorded dengue counts is modelled by the combination of an intercept term, a temporal effect and a seasonal effect: $f(t)=\iota+\alpha_t+\eta_t$. The temporal ($\alpha_t$) and seasonal ($\eta_t$) effects were defined using penalized cubic splines, and set up using the \texttt{jagam} function from the \texttt{mgcv} package for the R programming language \citep{jagam}. This was shown to be a very flexible model in capturing smooth temporal and seasonal variation, so we also consider it here to describe the time series of COVID-19 deaths counts for any individual region, though dropping the seasonal component (as we have only a few months of data). To capture spatio-temporal variability, we extend this to include spatially-varying intercept and temporal effects:
\begin{eqnarray}
f(t,s) = \iota_s + \delta_{t,s},
\end{eqnarray}
with $\iota_s$ assigned a non-informative Normal($0,10^2$) prior distribution and $\delta_{t,s}$ characterised as penalized cubic splines of time for each region, defined by $\delta_{t,s}=\bm{X}_t \bm{\kappa}_s^{(\delta)}$. Here $\bm{X}_t$ is a model matrix of the basis functions evaluated at each time point, and $\bm{\kappa}_s^{(\delta)}$ is a vector of coefficients. To penalize the splines for over-fitting, the coefficients are assigned a Multivariate-Normal prior with mean zero and precision matrix $\bm{\Omega}_s^{(\delta)} = \tau_s^{(\delta)} \bm{M}^{(\delta)}$. Matrix $\bm{M}^{(\delta)}$ is a known non-diagonal matrix, scaled by a smoothing (penalty) parameter $\tau_s^{(\delta)}$ \citep{jagam}, so that larger values of $\tau_s^{(\delta)}$ result in a smoother $\delta_{t,s}$ for each $s$.

As the regions in the data are geographically very large, we are more concerned with accounting for similarity in trends between regions -- in both the fatality rate and in the reporting delay over time -- than explicitly modelling any space-time interactions. To achieve this, we can re-introduce the temporal effect $\alpha_t$ and make its (basis function) coefficients the mean of the coefficients for the regional effects $\delta_{t,s}$, i.e.
\begin{eqnarray}
\alpha_t &=& \bm{X}_t \bm{\kappa}^{(\alpha)}\label{eq:overall}; \\
\bm{\kappa}^{(\alpha)} & \sim & \mbox{Multivariate-Normal}(\bm{0},\bm{\Omega}^{(\alpha)}=\tau^{(\alpha)}\bm{M}^{(\alpha)});\\
\bm{\kappa}_{s}^{(\delta)} & \sim & \mbox{Multivariate-Normal}(\bm{\kappa}^{(\alpha)},\bm{\Omega}_s^{(\delta)}) \label{eq:regional}.
\end{eqnarray}

The function $\alpha_t$ therefore captures common temporal variation across all regions (and so can be interpreted as the overall trend in the fatality rate for the whole of England), while the $\delta_{t,s}$ capture regional deviations from these overall trends. The parameter $\tau^{(\alpha)}$ therefore penalizes the overall (England) effect for smoothness, while the $\tau_s^{(\delta)}$ penalize the smoothness of the regional deviations from the overall effects. The main advantage of this structure -- which can be efficiently implemented using Markov Chain Monte Carlo (MCMC) owing to the conjugate relationship between the Multivariate-Normal priors for the overall effects and the regional deviations -- is that $\alpha_t$ can capture temporal covariation between regions, an important feature for some applications as discussed in Section \ref{sec:spatial}. This approach to pooling information, while allowing for individual variability, was shown to be very effective in modelling global trends in polluting cooking-fuel usage, so that countries with little data could borrow information from regional trends \citep{HAP}. 

We adopt the same approach when extending the relatively simple (Survivor) model used in \cite{DR} for the expected cumulative proportion reported at each delay, $g(t,d)=\psi_d + \beta_t$, first to include spatial variability and second to account for any weekly cycles (see Figure \ref{fig:new_deaths}) in the reporting delay:
\begin{eqnarray}
g(t,s,d) = \psi_{s,d} + \gamma_{t,s} +\xi_{t,s}.
\end{eqnarray}
Here fixed delay effects $\psi_{s,d}$ capture the overall curve of the cumulative proportion reported after each delay and are independent across regions. They are assigned non-informative first order random walk prior distributions, i.e. $\psi_{s,d} \sim \mbox{Normal}(\psi_{s,d-1},10^2)$, but truncated such that $\psi_{s,d}>\psi_{s,d-1}$ (to respect the fact that the cumulative proportion should increase with $d$). As for the model for $f(t,s)$, temporal effects $\gamma_{t,s}$ are penalized cubic splines centred on an overall temporal trend $\beta_t$ (as in \eqref{eq:overall}-\eqref{eq:regional}). Finally, $\xi_{t,s}$ are penalized cubic splines with a cyclic (periodic) basis over the days of the week - to account for systematic variability such as the `weekend-effect' - centred around an overall weekly effect $\eta_t$.

\subsection{Prior distributions and implementation}\label{sec:priors}
Prior distributions for other parameters were chosen to constrain the parameter space to reasonable values (relative to the data) but without being overly informative: For the Negative-Binomial dispersion parameters $\theta_s$ we specified independent Gamma(2,0.02) prior distributions, where the 95\% credible interval $[12.1,279]$ covers high levels of over-dispersion (e.g. $\theta_s=20$), while more extreme levels (e.g. $\theta_s=10$) are less likely a-priori.  We also specified Gamma(2,0.02) priors for the Beta-Binomial dispersion parameters $\phi_{s,d}$, following the same reasoning. Finally, it can be more interpretable to parametrize the spline precision penalties (e.g. $\tau_s^{(\delta)}$) as standard-deviation penalties (i.e. $\sigma_s^{(\delta)}=1/\sqrt \tau_s^{(\delta)}$), so that smaller values for $\sigma_s^{(\delta)}$ correspond to a stricter penalty. For these we specified positive Half-Normal(0,1) prior distributions, meaning smoother functions are more likely a-priori.

As discussed in \cite{DR}, instead of explicitly modelling all available partial counts $z_{t,s,d}$, we can reduce computational complexity by choosing to only explicitly model counts for $d\leq D'\leq D_{max}$. We achieve this by only including the conditional Beta-Binomial models for $z_{t,s,d}$ up to $D'$, so that the remainder $r_{t,s}=y_{t,s}-\sum_{d=D'+1}^D z_{t,s,d}$ is modelled implicitly. The trade-off associated with this choice is that predictive precision for $y_{t,s,d}$ is reduced, but generally only for past weeks $t\leq t_0 - D'$. Hence selecting a small $D'$ may be considered pragmatic where optimally precise predictions are not needed far into the past. In this experiment we opt for $D'=6$, which we consider sensible in a situation where optimally precise predictions are not needed for six weeks or more into the past. 

All code was written and executed in the R programming language \citep{R}. The model was implemented using the \texttt{nimble} package \citep{nimble}, which facilities highly flexible implementation of Bayesian models using MCMC. Four MCMC chains were run from different randomly generated initial values and with different random number generator seeds. We ran the chains for 200k iterations, discarding 100k as burn-in and then thinning by 10. Convergence of the MCMC chains was assessed by computing the potential scale reduction factor (PSRF) \citep{convergence} for samples of each $\lambda_{t,s}$, and of each $\theta_s$  (i.e. the parameters associated with nowcasting and forecasting). By convention, starting multiple chains from different initial values, with different random number generator seeds, and obtaining a PSRF close to or less than 1.05 for a given parameter is taken to indicate convergence. Here, all of the PSRFs for the $\theta_s$ were less than 1.05, as well as virtually all of the $\lambda_{t,s}$ ($>$93\%). The model was not fitted to the counts for the whole of England, but predictions for England are achieved by summation of Monte Carlo samples from each region's predictive distribution.

\subsection{Results for the 4th of May 2020}\label{sec:results}
To illustrate our approach as a tool for real-time decision-making, we first look at estimates and predictions from the model imagining we are fitting it at the end of the 4th of May 2020, with only data that would have been available then. In the following subsection, we then present a rolling prediction experiment to assess nowcasting and forecasting performance when the model is employed systematically over a longer period of time.
\begin{figure}[h!]
\includegraphics[width=\linewidth]{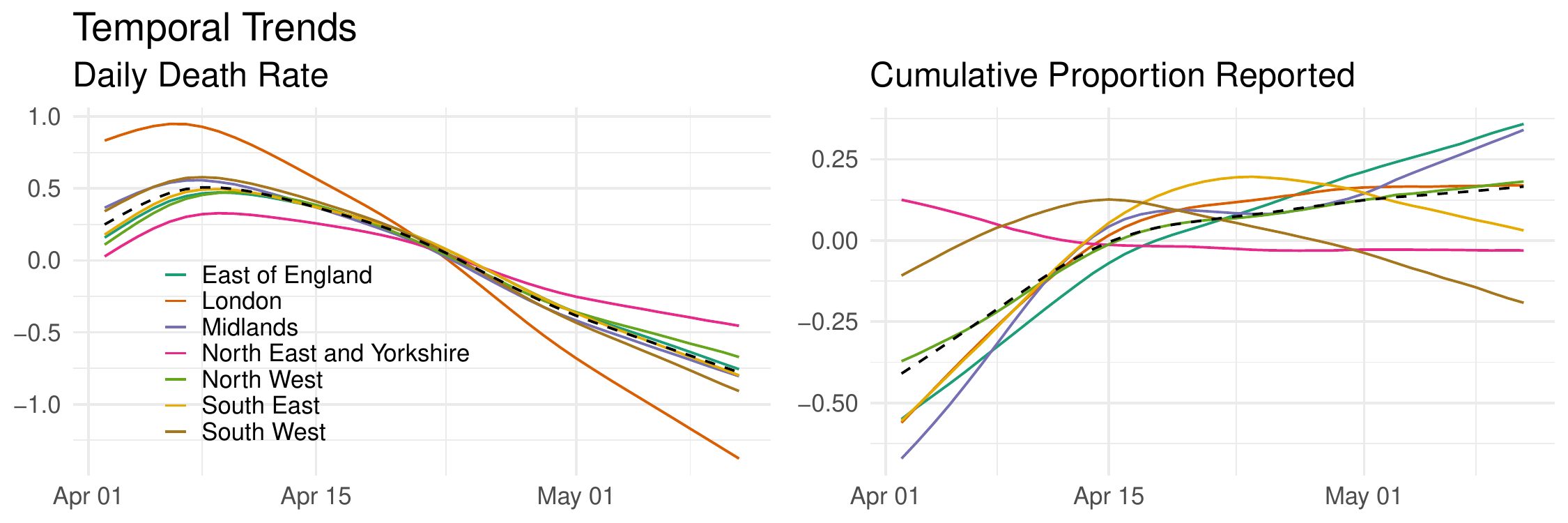}
\caption{Posterior median effects of time on the daily COVID-19 fatality rate (left) and the cumulative proportion reported (right), for each region. The dashed lines show the overall effects for England. }\label{fig:effects} 
\end{figure}

The left panel of Figure \ref{fig:effects} shows the posterior median temporal splines $\delta_{t,s}$ in the mean fatality rate $\lambda_{t,s}$. The dashed line shows the overall effect for England, $\alpha_t$. All regions show a peak around the first week of April, before decreasing at varying rates. Meanwhile, the right panel of Figure \ref{fig:effects} shows the posterior median temporal splines in the probit model for the cumulative proportion reported. Most regions (and indeed the overall effect for England) show an increase in the first two weeks of April, after which they are fairly constant, equating to a general speeding up of reporting in the East of England, London, the Midlands, the North West and the South East.
\begin{figure}[h!]
\floatbox[{\capbeside\thisfloatsetup{capbesideposition={right,center},
capbesidewidth=0.5\linewidth}}]{figure}[1\linewidth]
{\caption{Posterior median expected cumulative proportion of deaths reported after each day of delay, for each region.} \label{fig:delay_plot}}
{\includegraphics[width=\linewidth]{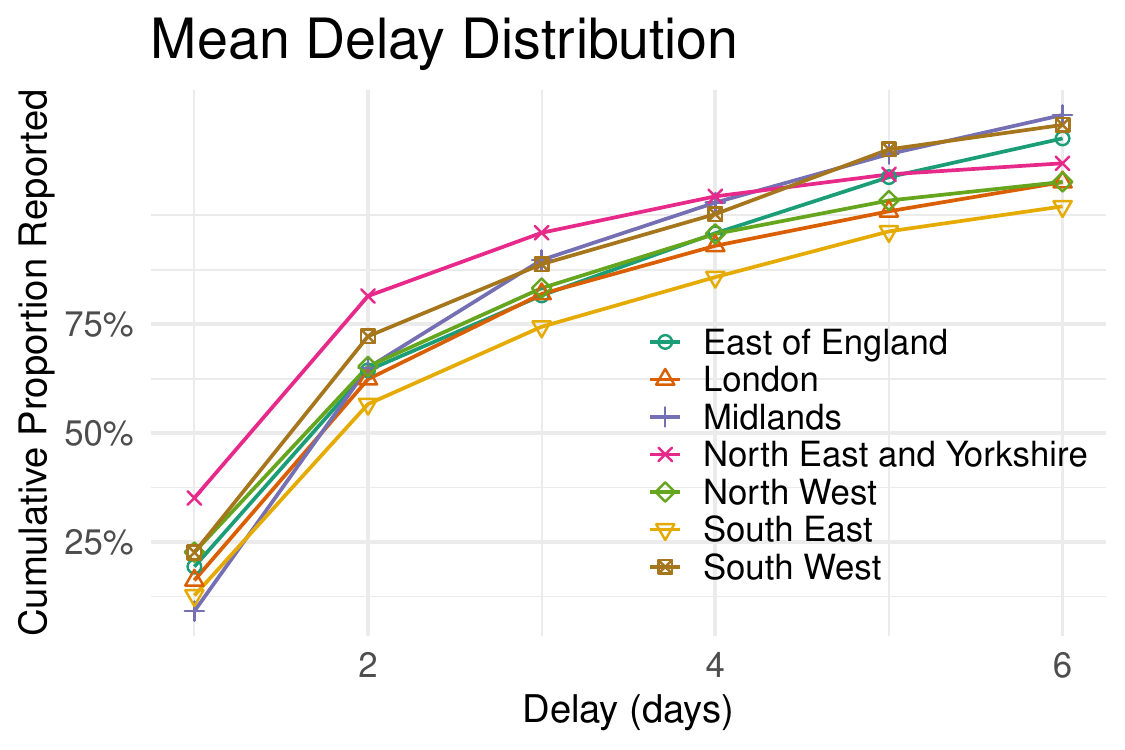}}
\end{figure}

Similarly, Figure \ref{fig:delay_plot} shows the posterior median of the overall mean delay distribution, as quantified by effects $\psi_{t,d}$ and plotted in terms of the cumulative proportion reported. We can interpret this plot as the expected delay distribution when temporal ($\gamma_{t,s}$) and weekly ($\xi_{t,s}$) effects are equal to zero. Recall from Section \ref{sec:nested} that these effects are estimated independently for each region. With that in mind, the delay curves are remarkably similar across regions, with around a quarter of deaths reported within the first reporting period and with most deaths reported within 2 days of occurrence.

\begin{figure}[h!]
\includegraphics[width=\linewidth]{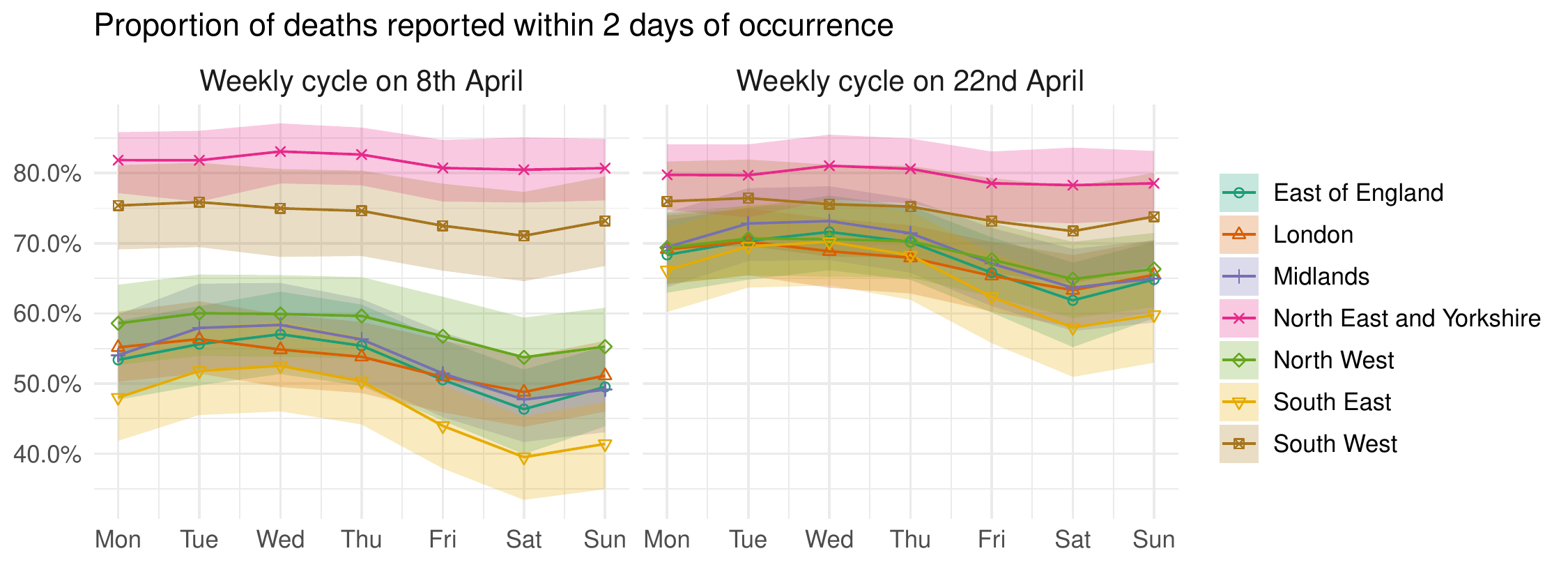}
\caption{Posterior median (with 95\% credible intervals) of the expected proportion of COVID-19 deaths reported within 2 days of occurrence ($S_{t,s,2}$), for each region and as a function of which day of the week they occurred. The temporal effects are fixed at their values for the 8th of April 2020 (left panel) and the 22nd of April 2020 (right panel). }\label{fig:weekly} 
\end{figure}

Combining all of the effects in Figures \ref{fig:effects} and \ref{fig:delay_plot} along with the weekly cycle ($\xi_{t,s}$), Figure \ref{fig:weekly} shows the posterior median expected proportion reported within 2 days ($S_{t,s,2}$) depending on which day of the week they occurred. Here, the temporal effects $\gamma_{t,s}$ have been frozen at their values on the 8th of April (left) and the 22nd of April (right). In both panels, we can see clear weekly cycles for most regions (the East of England, London,  the Midlands, the North West, and the South East), where a noticeably lower proportion of deaths occurring towards the end of the week are reported within 2 days. 

Looking across panels, the same regions exhibiting clear weekly cycles demonstrated a significant increase in the proportion reported within 2 days, from around 50\% to around 70\%, in just 2 weeks. This substantial change over a short period of just 14 days may not be captured well by approaches relying solely on a moving-window to account for systematic temporal variability in the delay distribution (i.e. \cite{nobs}), which has implications for nowcasting and forecasting accuracy, as we investigate in Section \ref{sec:compare}. The remaining two regions, the North East and Yorkshire and the South West, had consistently high reporting rates throughout the week and throughout the time period of study.
\begin{figure}[h!]
\includegraphics[width=\linewidth]{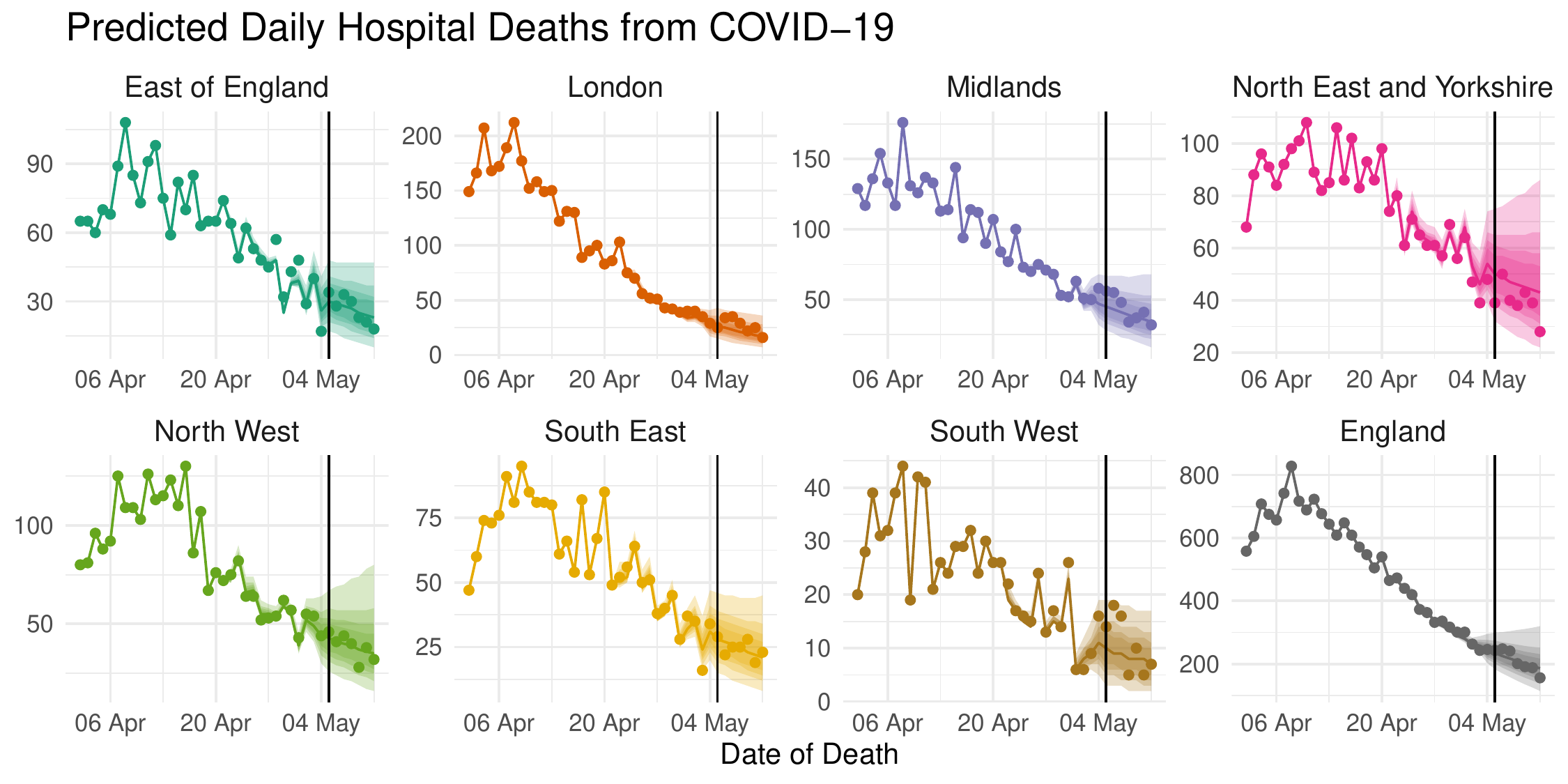}
\caption{Posterior median nowcasting and forecasting predictions of the total daily deaths $y_{t,s}$ (lines) with up to 95\% prediction intervals (shaded areas) for each region, using only data available on the 5th of May 2020 (vertical lines). Points show the total daily deaths reported within 14 days of occurrence (only available with hindsight). }\label{fig:forecast}
\end{figure}

Finally, Figure \ref{fig:forecast} shows nowcasting and forecasting predictions based on data available up to the 5th of May 2020. With hindsight, we can compare predictions to the now fully reported counts to assess performance, plotted as points. Generally the nowcasting predictions are good; forecasted trends are convincing, and uncertainty appears reasonable for decision-making purposes. To more thoroughly assess this, in the following subsection we present a rolling prediction experiment.

\subsection{Operational characteristics}\label{sec:operational}
To quantitatively assess nowcasting and forecasting performance when the model is systematically employed over a period of time, we carried out a rolling prediction experiment which emulates use of the model every day for 20 days. The experiment proceeds as follows:
\begin{description}
\item[Step 1:] Select a present-day $t_0$.
\item[Step 2:] Hold back all partial counts which would be unavailable at $t=t_0$.
\item[Step 3:] Fit the model, predicting any partially observed and completely unobserved deaths.
\item[Step 4:] Set $t_0=t_0+1$ and repeat steps 2-4.
\end{description}

\begin{figure}
\includegraphics[width=\linewidth]{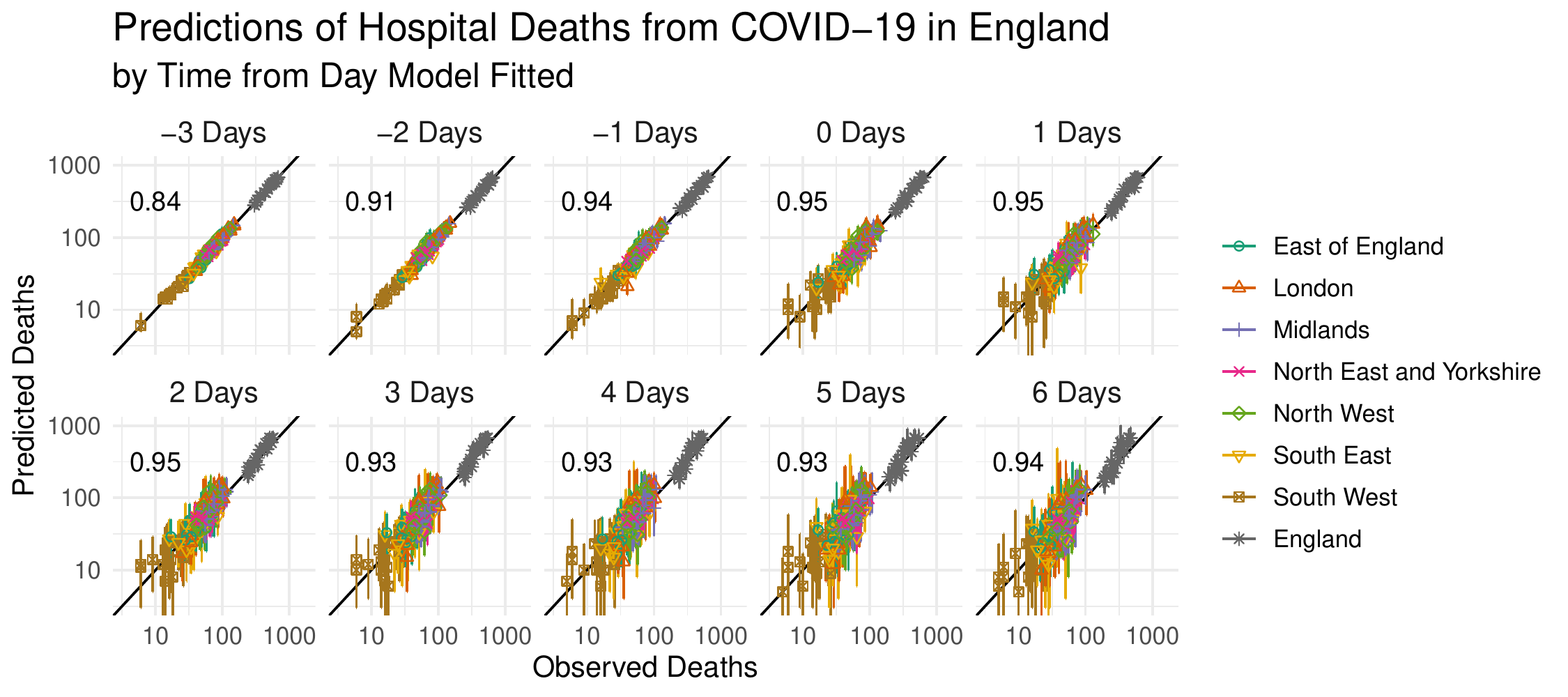}
\caption{Posterior median predictions (with 95\% prediction intervals) of daily hospital deaths in England, from the rolling prediction experiment, arranged by the time from the model fit day. Also shown are 95\% prediction interval coverage values.}\label{fig:coverage} 
\end{figure}

The prediction experiment starts on the 15th of April 2020, and ends on the 4th of May 2020, totalling 20 model runs. We then arrange predictions by the difference, in days, between the date the prediction is made for and the emulated date of fitting. For example, a difference of 0 days corresponds to all the predictions made for the same day the model was fitted. Moreover, a negative difference corresponds to a prediction made for a day in the past relative to the model fit (i.e. the count is at least partly observed) and a positive difference corresponds to forecasting. Figure \ref{fig:coverage} presents posterior median predictions of the total daily deaths (with 95\% prediction intervals) arranged in this manner. Also plotted are the 95\% prediction interval coverage values (the proportion of observations that lie within the 95\% prediction intervals). 

If the model is working as intended, we should see more accurate and less uncertain predictions for negative differences, because the counts have been more completely observed the further one predicts into the past and vice versa for forecasting. Meanwhile, we would hope to see high coverage values (90+\%) regardless of when we are making predictions for. First, looking at Figure \ref{fig:coverage} we do not see any evidence of systematic biases in the predictions (e.g. over- or under-predicting overall or for any particular region). The coverage values do appear to be a bit too low (meaning the model is over-confident) when making predictions a few days into the past, but the accuracy is so high that we doubt this would pose any practical problems. When nowcasting (i.e. a difference of 0 or perhaps -1 days) and when forecasting, the coverage is high, while the uncertainty is not unreasonable even when forecasting almost a week ahead in a fast-changing epidemic.

\subsection{Comparison with competing approaches}\label{sec:compare}
We have made a number of theoretical arguments in favour of the GDM over competing approaches (e.g. \cite{theo_dengue}, \cite{nobs}), particularly the separation of the four main sources of variability in disease surveillance data suffering from delayed reporting and the flexibility of the GDM as a model for random variability in the delay. However, it remains to be seen if these arguments translate into meaningful improvements in nowcasting performance when applied to real COVID-19 data in an operational setting. To investigate this, we seek to quantitatively compare 5 competing models -- based on the rolling prediction experiment (detailed in Section \ref{sec:operational}), focussing on same-day nowcasting predictions -- defined as predictions for the present day in a scenario where the model is fitted each day using the latest available data. Full details for each competitor model are provided in Appendix B, but they can be summarised as follows:
\begin{description}
\item[1) GDM] The full GDM as described in Section  \ref{sec:nested}.
\item[2) NB] A Negative-Binomial approximation to the marginal model of the GDM for the delayed counts $z$, where the models for systematic variability in the total count and in the delay are appropriately separated and are identical to those in Section \ref{sec:nested}.
\item[3) INLA] A ``direct'' Negative-Binomial model for the delayed counts $z$, based on the framework proposed by \cite{theo_dengue} and where the seasonal component is replaced by a weekly cycle to capture systematic weekly variability in the delay.
\item[4) NobBS] A model for the delayed counts $z$ based on the framework proposed in \cite{nobs} and implemented using the \texttt{NobBS} package for R. 
\item[5) NobBS-14] A second model based on \cite{nobs}, where a moving window of 14 days is specified to capture systematic temporal variation in the reporting delay.
\end{description}
Including a `marginal' version our GDM model (NB) based on the same systematic model for mean total count (deaths) and the same `Survivor' model for the expected cumulative proportion reported at each day should illuminate to what degree differences between the GDM and the other models (i.e. INLA\ and NobBS) are attributable solely to the use of the full GDM conditional model to appropriately capture variability in the reporting delay. So that the comparison can focus primarily on the performance of each modelling framework, rather than any specific spatio-temporal structures, all models are implemented as independent time series models for each of the 7 regions.
\begin{table}[h!]
\footnotesize
\begin{tabular}{llllllllllll}
\hline
\textbf{Region}          & \multicolumn{5}{c}{\textbf{$\sqrt{\mbox{Mean Squared Error}}$}} & \multicolumn{5}{c}{\textbf{Bias}}  \\
                         & GDM         & NB          & INLA        & NobBS       & NobBS-14    & GDM           & NB            & INLA          & NobBS         & NobBS-14      \\ \hline
East of England          & 10          & 9           & 14          & 15          & 14          & 3             & 3             & 10            & 13            & 11            \\
London                   & 18          & 17          & 28          & 32          & 30          & 7             & 7             & 18            & 25            & 20            \\
Midlands                 & 13          & 12          & 17          & 20          & 21          & 2             & 2             & 10            & 15            & 13            \\
N.E. and Yorks. & 8           & 8           & 8           & 7           & 8           & 1             & 2             & 5             & 5             & 5             \\
North West               & 17          & 16          & 18          & 19          & 19          & 6             & 6             & 11            & 14            & 11            \\
South East               & 13          & 12          & 16          & 20          & 18          & -1            & 2             & 11            & 16            & 12            \\
South West               & 6           & 5           & 5           & 5           & 5           & -1            & 0             & 1             & 2             & 1             \\ \hline
\textbf{Overall}         & \textbf{13} & \textbf{12} & \textbf{17} & \textbf{19} & \textbf{18} & \textbf{2}    & \textbf{3}    & \textbf{10}   & \textbf{13}   & \textbf{10}   \\\hline
 & \multicolumn{5}{c}{\textbf{Mean 95\% Prediction Interval Width}} &  \multicolumn{5}{c}{\textbf{95\% Prediction Interval Coverage}} \\
                         & GDM         & NB          & INLA        & NobBS       & NobBS-14    & GDM           & NB            & INLA          & NobBS         & NobBS-14      \\ \hline
East of England          & 46          & 52          & 67          & 69          & 62          & 1             & 1             & 1             & 0.95          & 0.95          \\
London                   & 59          & 61          & 99          & 108         & 88          & 0.9           & 0.95          & 0.95          & 0.95          & 0.95          \\
Midlands                 & 62          & 79          & 123         & 133         & 105         & 1             & 1             & 1             & 1             & 1             \\
N.E. and Yorks. & 44          & 53          & 61          & 56          & 60          & 1             & 1             & 1             & 1             & 1             \\
North West               & 58          & 77          & 100         & 102         & 81          & 0.9           & 1             & 1             & 1             & 1             \\
South East               & 45          & 60          & 78          & 87          & 70          & 0.9           & 1             & 1             & 1             & 0.95          \\
South West               & 21          & 22          & 27          & 25          & 24          & 1             & 1             & 0.95          & 0.95          & 0.95          \\
\hline \textbf{Overall}         & \textbf{48} & \textbf{58} & \textbf{79} & \textbf{83} & \textbf{70} & \textbf{0.96} & \textbf{0.99} & \textbf{0.99} & \textbf{0.98} & \textbf{0.97} \\ \hline
\end{tabular}
\caption{Nowcasting performance of competing models when applied to COVID-19 data.}\label{tab:compare}
\end{table}

To quantify differences in nowcasting performance, we calculate 4 metrics for each model. The first is the root-mean squared error of the posterior median predicted number of deaths on each day, which quantifies how accurate point estimates are. The second is the bias, defined as the mean difference between the median predicted deaths and the observed deaths, which quantifies any systematic over- or under-prediction when nowcasting. The third is the mean 95\% prediction interval width for the total number of deaths on each day, which quantifies how precise/uncertain predictions are. The fourth and final metric is the 95\% prediction interval coverage, which checks whether uncertainty is reliably quantified by the model. In particular, coverage values much less than 0.95 might suggest too few data points are captured by the 95\% intervals and the model is {\it over-confident}. Conversely, higher coverage values could suggest the predictions display excessive uncertainty. 
\begin{figure}[h!]
\includegraphics[width=\linewidth]{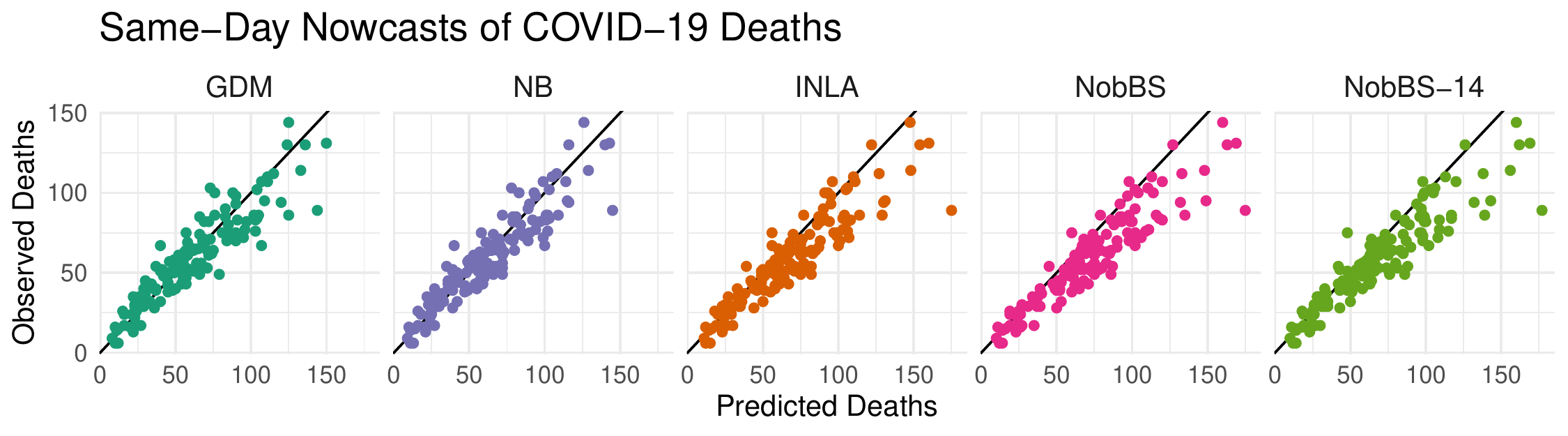}
\caption{Scatter plots of posterior median predicted total deaths (across all regions) versus observed total deaths, from each competing model.}\label{fig:scatter} 
\end{figure}

Table \ref{tab:compare} presents these metrics separately for each region and overall. Note that the rolling prediction experiment covers 20 days, therefore resulting in 20 same-day predictions for each region. This means a difference in a regional coverage value of 0.05 equates to one data point being captured or not captured. In every region predictions from the GDM were the most precise (as quantified by lower prediction interval widths), while point estimates from the GDM and the marginal NB (which share the same systematic model) were the most accurate in 5 out of 7 regions and equally accurate in the remaining 2. Predictions from the GDM and the marginal NB did not display the persistent over-prediction seen in the INLA and NobBS approaches -- illustrated in Figure \ref{fig:scatter} which compares posterior median predicted deaths with observed deaths -- which we believe is likely due to the use of splines to capture the downward trends in the total deaths rather than the first order random walk (RW1) models in the INLA and NobBS models. Overall, the GDM outperformed the INLA (based on \cite{theo_dengue}) and NobBS models (based on \cite{nobs}) by wide margins across the board. All models have overall coverage values above 0.95, with the GDM offering the closest at 0.96.
\begin{figure}
\includegraphics[width=\linewidth]{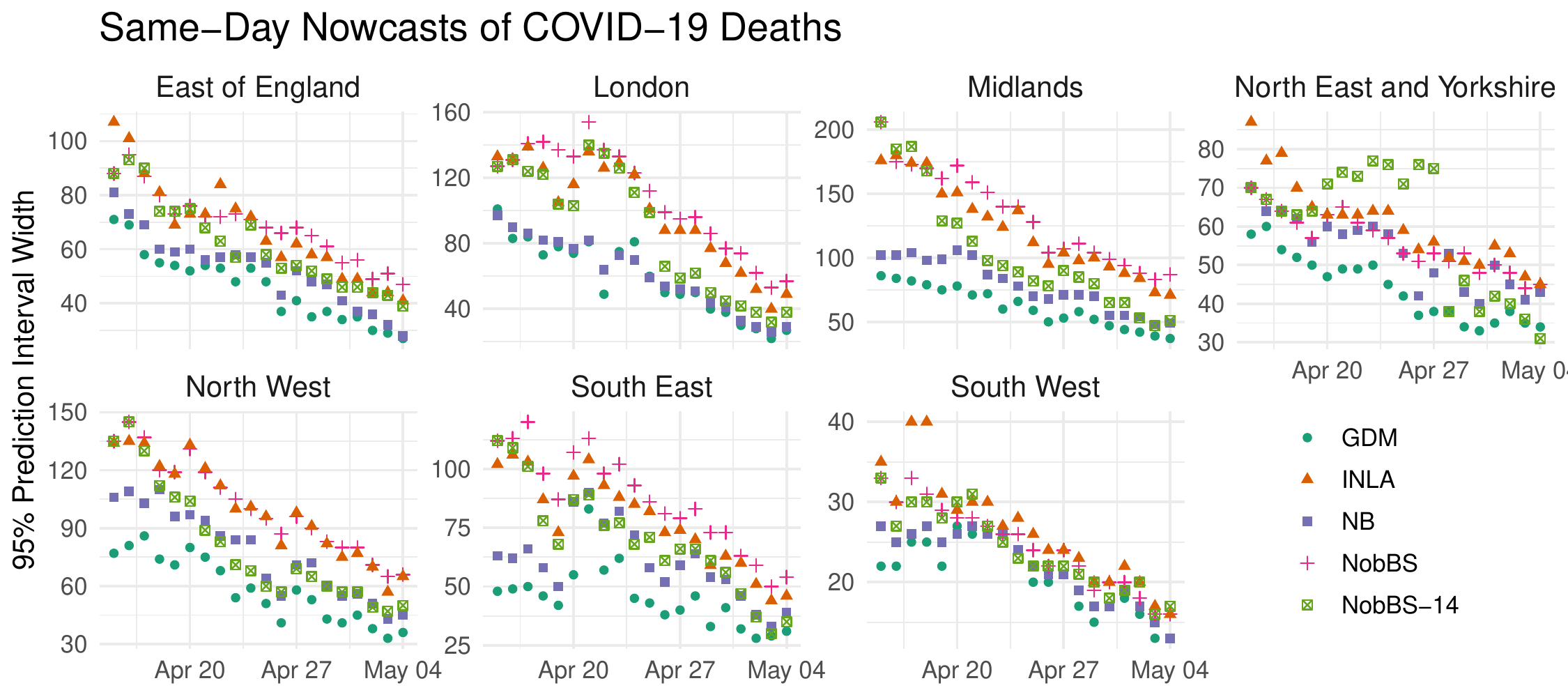}
\caption{Time series of central 95\% prediction interval widths for same-day nowcasts from the 5 competing models, for each region.}\label{fig:piw} 
\end{figure}

Figure \ref{fig:piw} shows 95\% prediction interval widths as a time series for each day. Here, consistent leads in the precision of nowcasting predictions from the GDM compared to competing frameworks (INLA, NobBS) are stark, particularly earlier on in the pandemic. Meanwhile, the marginal (NB) approximation to our GDM model also outperforms a) the INLA model, likely due to the appropriate separation of systematic variability in the total count and in the delay; and b) the NobBS models, probably due to the inclusion of a weekly cycle and due to splines being a more flexible way of capturing systematic temporal variation in the delay compared to a fixed moving window. However, the importance of appropriately quantifying delay variability using a flexible model like the GDM is illustrated by the non-trivial difference between the marginal model (NB) and the GDM -- noting that the marginal (NB) model still relies on a full MCMC implementation due to the non-linearity of explicitly modelling the cumulative proportion reported at each delay.

All code and data required to reproduce these results are provided as supplementary material.

\section{Discussion}\label{sec:discuss}
The COVID-19 pandemic has highlighted the need to optimally correct delays in disease data, for timely mitigation actions. Here, we have critically reviewed the three mainstream approaches to correcting delays, and quantified their respective performance when applied to COVID-19 mortality data. We have argued that our multivariate approach based on the Generalized-Dirichlet-Multinomial is theoretically the most advanced in explicitly capturing the different sources of variability in the data. Furthermore, of the three it is the only one which readily provides predictions of total counts $y$ conditional upon {\it all} available data, i.e. historic $y$ and partial counts $\bm{z}$. Indeed, when applying the GDM and four other competing models representing the current best-practice in addressing the delayed reporting to the COVID-19 deaths data, the GDM approach is by far the most capable in terms of nowcasting accuracy, bias, and precision -- in a realistic operational scenario. Compared to competitors, the GDM exhibited around a 25-30\% smaller root-mean squared error and around a 30-40\% smaller mean 95\% prediction interval width, whilst still offering 95\% prediction interval coverage values in the region of 0.95. This certainly establishes the GDM as the current best-practice in nowcasting COVID-19 and other diseases, while also allowing for reliable forecasting.

To account for spatio-temporal variability, we have introduced a novel spatial extension to the GDM framework -- allowing for a wide variety of spatial, temporal, and spatio-temporal structures to be included in both the model for the total reported counts after any delays have passed, and in the model for the delay mechanism itself. Within this we have demonstrated one specific model based on nested spline structures. Though the spatio-temporal models presented here for COVID-19 and SARI data are quite intuitive and demonstrably better than independent time series models when aggregating predictions to a super-regional level (as illustrated in Appendix A.3), they may be overly simplistic: Firstly, while our nested spline approach enabled the model to capture the distribution of the total counts across all regions well, both the model for the total deaths/cases and the delay model lack any explicit spatial structure (i.e. the model assumes equivalent similarity between all regions). A more sophisticated approach to space-time interactions may be needed for applications with a fine spatial resolution, e.g. to potentially capture the spread of a disease over time. Furthermore, in cases where some regions have a lot of missing data, models with explicit spatial structure may allow for more precise predictions in those regions. A lack of data may also warrant including spatial structure in the dispersion parameters, which is possible by modelling them as log-linear \citep{DR}. Applications intended for operational use might also benefit from considering more complicated mean delay models with delay-time interactions, which are of course possible within the framework proposed here, e.g. using tensor product smooths \citep{tensor}. 

The model we presented for correcting delayed of COVID-19 deaths takes around 10-20 minutes to compile and run (as implemented in Section \ref{sec:priors}). We believe this is reasonable in a daily operational setting, allowing for potential errors and any need to run the MCMC for more iterations for convergence. Indeed, a model based on \cite{DR} for nowcasting daily COVID-19 deaths by age and region in England \citep{Seaman} has been implemented operationally and currently feeds into the UK Scientific Pandemic Influenza Group on Modelling (SPI-M) on a weekly basis \citep{mrc_bsu}. However, the greater dimensions of the SARI data (more time points and spatial regions) results in a run-time in the region of 12 hours. \cite{Seaman} improved on our implementation of the GDM by not sampling any missing partial counts $z$, leading to better mixing of MCMC chains, which would go some way in reducing overall run-times. However, for applications with even greater dimensions (e.g. COVID-19 data at the level of individual cities, hospitals, or trusts), there is a clear need for either a method of implementing the GDM in a more efficient manner or a new framework altogether which offers comparable predictive performance to the GDM and improved computational feasibility.

Using the framework presented here it is possible to comprehensively address the problem of delayed reporting in COVID-19 and other disease surveillance applications. This ignores, however, the possible problem of substantial under-reporting in the final reported counts, e.g. due to a lack of testing capacity, as described in Appendix C. Therefore any predictions from models discussed here, e.g. of disease cases, may lead to an undersized response to the true magnitude of an outbreak. Although we have explained how under-reporting can be taken into account within the modelling framework we propose, this approach has so far only been applied in a purely hypothetical scenario (see \cite{thesis}). Further research should therefore be directed at applications where effective disease surveillance is inhibited by both delayed reporting and under-reporting.

\bigskip
\begin{center}
{\large\bf SUPPLEMENTARY MATERIAL}
\end{center}
All supplementary files are contained within an archive which is available
to download as a single file.
\begin{description}

\item[COVID Master:] Master script for reproducing the COVID-19 application (R script).

\item[COVID Data:] Data for the COVID-19 application (Excel spreadsheet).
\item[SARI Master:] Master script for reproducing the SARI application (R script).
\item[SARI Data:] File containing data for the SARI application (RData file).
\end{description}

\bibliographystyle{Chicago}

\bibliography{library}

\begin{appendices}

\section{Severe Acute Respiratory Infection Data}\label{app:sari}
The World Health Organization defines a severe acute respiratory infection (SARI) as an acute respiratory infection where the patient suffers from both a fever measured above 38$^{\circ}$C and coughing, where hospitalization is necessary and where the onset of the infection was within the last 10 days \citep{sari_definition}. One reason for this classification is to standardise surveillance of influenza-like illnesses, so that seasonal patterns in respiratory virus circulation can be studied and to inform prevention policies \citep{theo_dengue}. As explained in \cite{theo_dengue}, lags in data assimilation (from hospitals to local authorities, then to state and national levels) introduce delays in the information on SARI available to public health decision-makers, potentially inhibiting response to influenza outbreaks. In this section we apply our proposed framework to SARI data from Brazil, to illustrate the generality of our approach beyond COVID-19, as well as to compare a joint (spatio-temporal) model to independent time series models.

\subsection{Data}
We use data from the Brazilian state of Paran\'{a}, which was severely affected by the 2009 H1N1 epidemic compared to other states \citep{H1N1} and continues to have one of the highest rates of SARI incidence \citep{theo_dengue}. Here we consider a much longer time period of 230 weeks (from the start of January 2013 to the end of May 2017), compared to 66 weeks in \cite{theo_dengue}, to enable us to draw meaningful conclusions about seasonal variation. The state is divided into $s=\{1,\dots,22\}$ health regions and we consider the total count to be fully observed 6 months after occurrence ($D_{max}=27$). The dimension of the total counts $y_{t,s}$ is therefore 230x22 and the dimension of the partial counts $z_{t,s,d}$ is 230x22x27 (corresponding to over 100k observations). For this application, we imagine that the present-day week, $t_0$, is week 224 (mid April 2017). We then seek to make predictions for $t_0=224$, for previous weeks where the total count is still partially unobserved ($t=\{t_0-D_{max}+2,\dots,t_0-1\}=\{199,\dots,223\}$) and for the next 6 weeks ($t=\{t_0+1,\dots,t_0+6\}=\{225,\dots,230\}$).
\begin{figure}[h!]
\includegraphics[width=\linewidth]{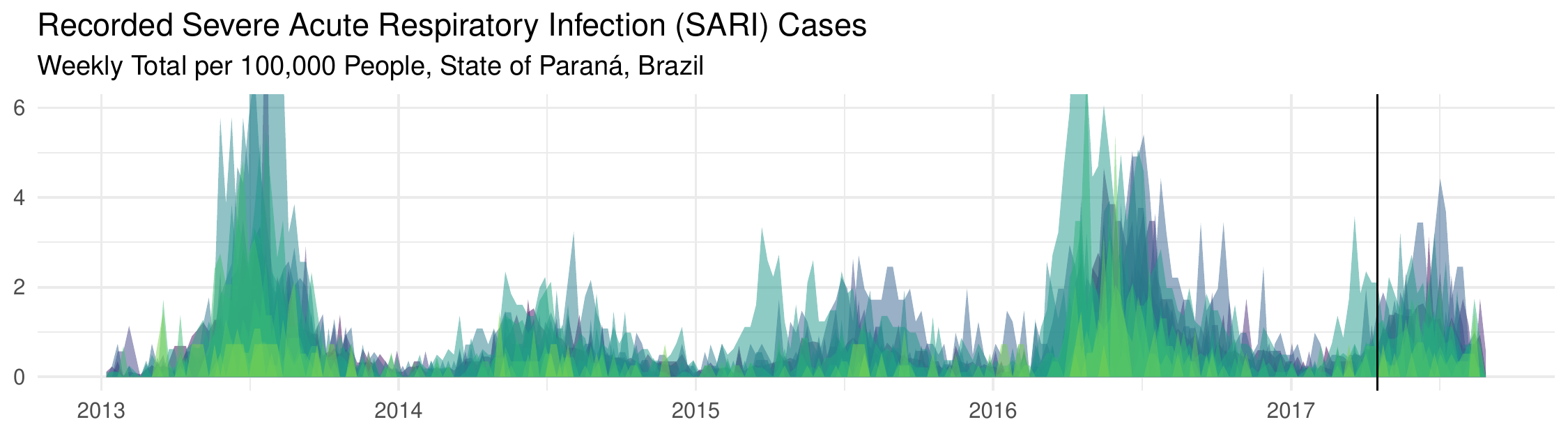}
\caption{Area plot of total recorded SARI cases per 100,000 people, with a different colour for each of the 22 health regions. The vertical line shows the present week for this experiment $t_0=224$.}\label{fig:data} 
\end{figure}

Figure \ref{fig:data} shows weekly total recorded SARI cases per 100,000 people by region. The plot shows a clear seasonal cycle across all regions, with outbreaks reaching their peak in April-July, as well as considerable year-to-year variability. There is also some evidence of regional variation in the overall rate -- for example, the brightest green region tends to have quite a low rate of cases per 100,000 people, compared to some other regions -- as well as regional variation in the seasonal timing of outbreaks. At ``present day'' $t_0=224$, shown by the vertical line, we are in the early stages of the annual influenza outbreak, so forecasting predictions should ideally show an increasing trend in the number of SARI cases.

\subsection{Model for SARI cases}
In Section 4.1, we used a nested spline structure to add a spatial dimension to the time series model for dengue fever cases presented in \cite{DR}. To model the incidence of SARI cases we adopt a near identical approach, making use of spatially-varying intercept, temporal and seasonal effects:
\begin{eqnarray}
f(t,s) = \iota_s + \delta_{t,s} + \xi_{t,s};
\end{eqnarray}
Effects $\delta_{t,s}$ and $\xi_{t,s}$ are temporal and seasonal (cyclic) splines, respectively, which vary by region. Figure \ref{fig:data}, however, suggests that a large portion of temporal and seasonal variation may be common across all regions. Once again, we can introduce temporal and seasonal effects $\alpha_t$ and $\eta_t$, and make their (basis function) coefficients the mean of the coefficients for the regional effects $\delta_{t,s}$ and $\xi_{t,s}$. Similarly, the model for the expected cumulative proportion reported after each delay is characterised by the addition of fixed delay effects $\psi_{s,d}$ which are independent across regions and temporal spline effects $\gamma_{t,s}$:
\begin{eqnarray}
g(t,s,d) = \psi_{s,d} + \gamma_{t,s}.
\end{eqnarray}
Prior distributions and implementation are the equivalent to the COVID-19 model, detailed in Section 4.2.

\subsection{Results}\label{app:sari:results}
\begin{figure}[h!]
\includegraphics[width=\linewidth]{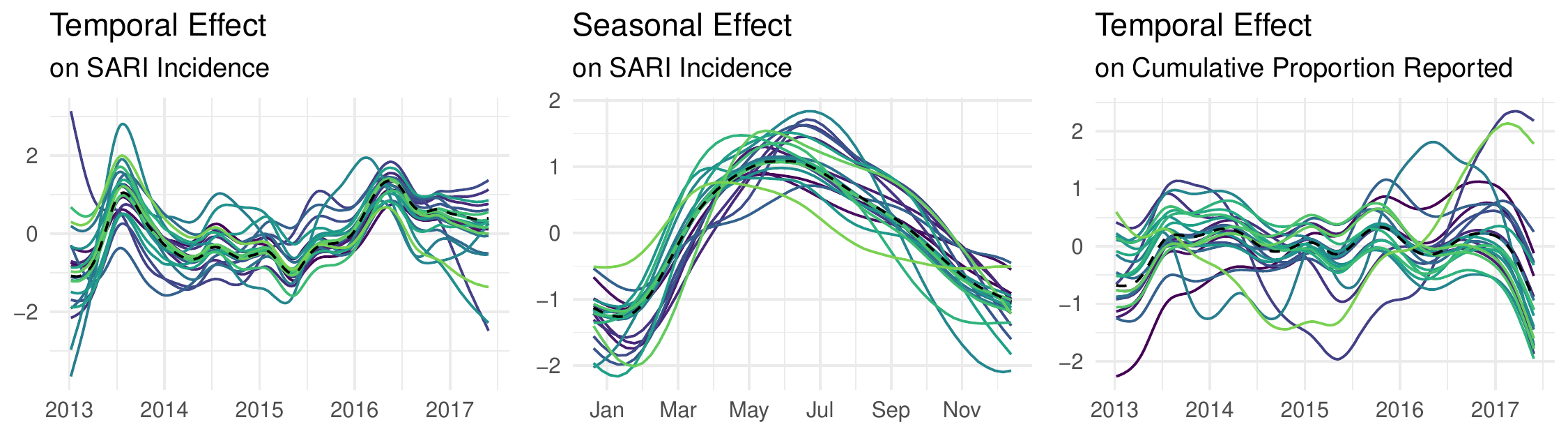}
\caption{Median predicted temporal and seasonal effects on SARI incidence (left and centre) and median predicted temporal effect on the cumulative proportion reported (right).\label{fig:sari:effects}}
\end{figure}

Figure \ref{fig:sari:effects} shows median predicted temporal ($\delta_{t,s}$, left) and seasonal ($\xi_{t,s}$, centre) effects on SARI incidence, as well as the temporal effect on the cumulative proportion reported ($\gamma_{t,s}$, right). A different colour is used  for each region and the dashed black lines show the median predicted overall effects, $\alpha_t$, $\eta_t$ and $\beta_t$, respectively. The estimated effects on SARI incidence follow the overall trends quite closely, with only a few deviating substantially. For example, there are noticeable increases in the temporal effect on SARI incidence for almost all regions around mid 2013 and around mid 2016, corresponding to the two largest outbreaks seen in Figure \ref{fig:data}. Similarly, all the seasonal effects reflect the increase in SARI incidence leading up to Brazil's winter seen in Figure \ref{fig:data}. The effects on the cumulative proportion reported are substantially more variable, suggesting that the delay mechanism may be driven more by local factors compared to SARI incidence.

Summarising, the 22 health regions of Paran\'{a} have a lot in common, in terms of temporal and seasonal variation in SARI incidence. It is worth examining, however, whether anything tangible was gained from modelling the regions simultaneously as opposed to using 22 independent time series models. In  Section 3.1, we argued that modelling the regions independently could impede the model's ability to capture the variance of the total number of reported cases across all regions. To assess this, we applied 22 independent models where $f(t)=\iota + \delta_{t} + \xi_{t}$ and $g(t,d)=\psi_{d}+\gamma_t$ for each region. We then used posterior predictive checking \citep{Gelman2013} to see which approach captures the variance of the total better.
\begin{figure}[h!]
\floatbox[{\capbeside\thisfloatsetup{capbesideposition={right,center},capbesidewidth=0.48\linewidth}}]{figure}[1\linewidth]
{\caption{Posterior replicates of the sample mean (left) and sample variance (right) of fully observed ($t\leq t_0-D_{max}+1$) total reported number of SARI cases across all 22 health regions. Vertical lines show the corresponding observed statistics.} \label{fig:replicate_plot}}
{\includegraphics[width=\linewidth]{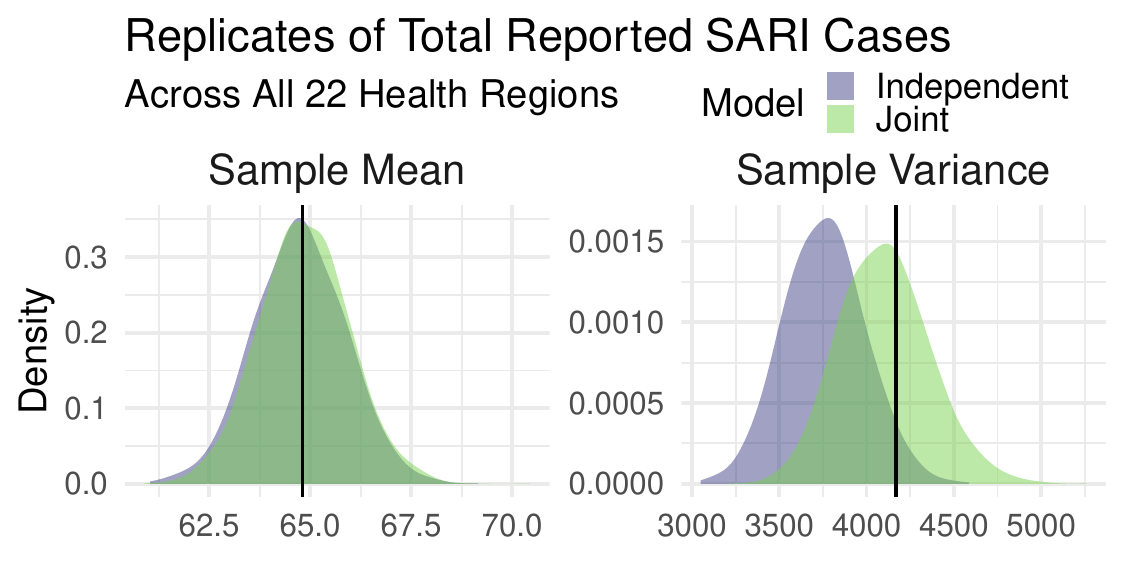}}
\end{figure}

Figure \ref{fig:replicate_plot} shows posterior replicates of the sample mean and sample variance of fully observed ($t\leq t_0-D_{max}+1$) total reported number of SARI cases across all 22 health regions (i.e. the total for the whole state). Both models perform equally well at capturing the mean number of cases, but notably the joint model is overwhelmingly better at capturing the sample variance compared to the independent models -- the observed sample variance is firmly in the centre of the replicate distribution from the joint model. This suggests that the joint model is more appropriate in applications where predictions are also important at a larger geographical scale.

\begin{figure}[h!]
\includegraphics[width=\linewidth]{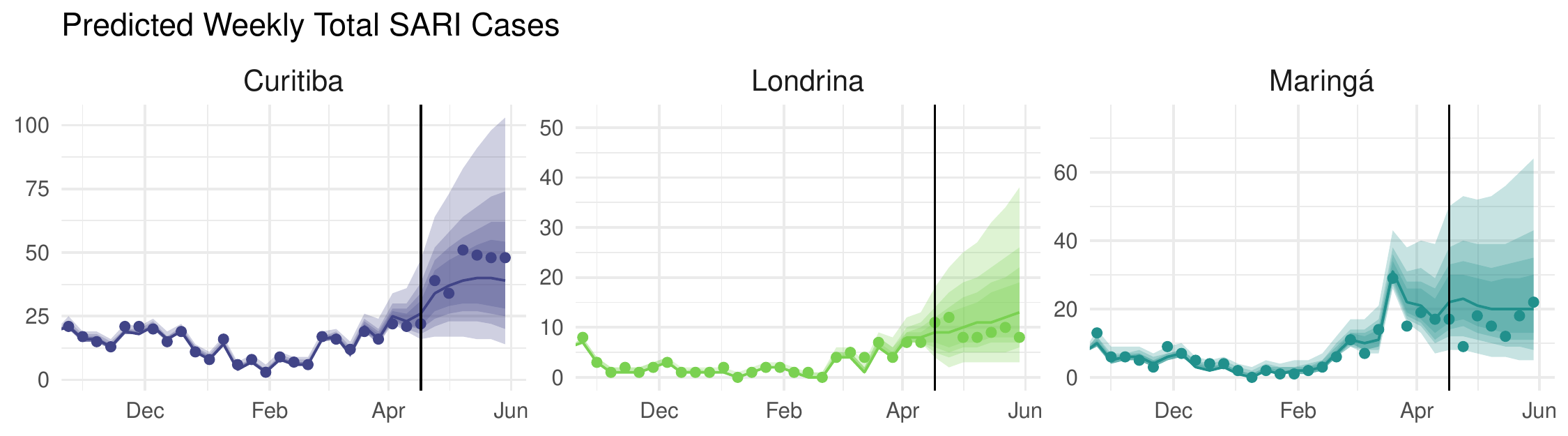}
\caption{Predicted (median, 50\%, 65\%, 80\%, and 95\% prediction intervals) total reported SARI cases for the three most populous health regions. The vertical lines show the present week $t_0=224$.}\label{fig:predict}
\end{figure}

Finally, we can examine the model's ability to nowcast and forecast in this application. Figure \ref{fig:predict} shows predicted total reported SARI cases in the three most populous regions of Paran\'{a}: Curitiba (left), Londrina (centre) and Maring\'{a} (right). Among these three regions, the forecasting predictions when fitted at present week $t_0=224$ appear most precise for Curitiba and Londrina, while somewhat over-predicting the number of cases in Maring\'{a}. This over-prediction is common across many of the regions, likely owing to the reduced magnitude of the outbreak compared to the previous year, which the model may detect when fitted further into the outbreak. That said, virtually all of the observations are within the 95\% prediction intervals, with precise nowcasting predictions for the present week (shown by the vertical line), which are all within the 50\% prediction intervals. 

For this experiment, the overall 95\% prediction interval coverage was 0.99 for predictions of the total reported count corresponding to previous weeks ($t<t_0$), 1 when forecasting ($t>t_0$) and 1 when nowcasting ($t=t_0$).

All code and data required to reproduce these results are provided as supplementary material.

\section{Details of Competing Models for COVID-19 Deaths}\label{app:compare}
In Section 4.5 we compared the GDM against four competing models for COVID-19 deaths in a rolling nowcasting experiment. Here we provide details on the specification and implementation of each of these models.

\subsection{Marginal Negative-Binomial Model (NB)}
The first competing model is a Negative-Binomial model intended to approximate the (as-yet) unknown marginal model for the delayed counts $z_{t,s,d}$ obtained from the GDM when integrating out the total counts $y_{t,s}$. The key feature of this model is that the systematic models for the total count and the reporting delay are otherwise identical to those presented in Section 4.1:
\begin{eqnarray}
z_{t,s,d}\mid \mu_{t,s,d},\lambda_{t,s}, \theta_s &\sim & \mbox{Negative-Binomial}(\mu_{t,s,d}\lambda_{t,s},\theta_s);\\
\log(\lambda_{t,s}) &= & \iota_s + \delta_{t,s};
\end{eqnarray}
where $\mu_{t,s,d}$ is the mean proportion at each delay, again defined by a probit model for the cumulative proportion reported:
\begin{eqnarray}
\mu_{t,s,d}=S_{t,s,d}-S_{t,s,d-1};\\
\mbox{probit}(S_{t,s,d})=\psi_{s,d} + \gamma_{t,s} +\xi_{t,s},
\end{eqnarray}
and $\lambda_{t,s}$ is the mean of $y_{t,s}$. Like in the GDM model for COVID-19 deaths, we include models for the first $D'=6$ delayed counts $z_{t,s,d}$, but also an extra (identical) model for the remainder $r_{t,s}=y_{t,s}-\sum_{d=1}^{D'} z_{t,s,d}$ -- which is unobserved when $y_{t,s}$ is unobserved (see \cite{DR} for the rationale behind modelling the remainder). The model is implemented using MCMC following the specification in Section 4.2 for prior distributions and implementation.

\subsection{Negative-Binomial Model Based on Bastos et al. (INLA)}
The second model is a Negative-Binomial model for the delay counts $z_{t,s,d}$ based on the framework and implementation detailed in \cite{theo_dengue}:
\begin{eqnarray}
z_{t,s,d}\mid \lambda_{t,s}, \theta_s &\sim & \mbox{Negative-Binomial}(\lambda_{t,s},\theta_s);\\
\log(\lambda_{t,s}) &= & \iota_s + \delta_{t,s} + \beta_{d,s}  +  \gamma_{t,s,d} + \xi_{t,s}.
\end{eqnarray}
Independently for each region, $\iota_s$ are the intercepts, $\delta_{t,s}$ are first order random walks (RW1) (to capture the epidemic curve), $\beta_{d,s}$ are RW1 effects to capture the mean delay distribution, $\gamma_{t,s,d}$ are RW1 temporal effects for each delay to capture temporal variation in the delay distribution, and $\xi_{t,s}$ are cyclic RW2 effects of time of day, to capture weekly cycles in the reporting delay. The main difference between this model and the above marginal approximation to the GDM is the  lack of separation of systematic variability in the total count and the reporting delay. The model covers each of the $d=1,\dots,14$ delays and is implemented using the Integrated Nested Laplace Approximation method and the \texttt{r-inla} package \citep{r-inla}, using code adapted from \cite{theo_dengue}.
\subsection{Models based on McGough et al. (NobBS and NobBS-14)}
The third and fourth competitor models are based on the framework for the delayed counts $z_{t,s,d}$ proposed by \cite{nobs} and implemented using the \texttt{NobBS} package for R. The package allows for both Poisson and Negative-Binomial models, and we opted for the latter to account for over-dispersion:
\begin{eqnarray}
z_{t,s,d}\mid \lambda_{t,s,d}, \theta_s &\sim & \mbox{Negative-Binomial}(\lambda_{t,s,d},\theta_s);\\
\log(\lambda_{t,s,d}) &= & \alpha_{t,s} + \log(\beta_{d,s});
\end{eqnarray}
where $\alpha_{t,s}$ is an RW1 effect to capture the epidemic curve and $\beta_{d,s}$ are vectors of proportions such that $\sum_{d=1}^{D_{max}} \beta_{d,s}=1$ (recall the maximum delay $D_{max}=14$). Effects $\beta_{d,s}$ therefore capture the expected proportion reported at each delay and are modelled as fixed in time. To account for temporal heterogeneity in the delay distribution, \cite{nobs} propose fitting the model to data in a moving window of fixed length, so that the estimated $\beta_{d,s}$ are more appropriate for recent data. We fit two models using this approach, one with no window (NobBS) -- the model is fitted to all of the data -- and one with a window of 14 days (NobBS-14), so that the estimated mean delay distribution is representative of the last two weeks. In both cases, we used the default weakly-informative prior distributions from the \texttt{NobBS} package, which implements the model using the JAGS software facility for MCMC \citep{JAGS}.

\section{Under-reporting}\label{app:ur}
Where count data are affected by delayed-reporting, the total reported count, $y_{t,s,d}$, is often still a substantial under-representation of the true count, termed here $x_{t,s,d}$. Reports of COVID-19 cases may, for instance, be affected by under-reporting due to a lack of testing availability, false negative test results, or a lack of symptoms. Similarly, some deaths due to COVID-19 may be missed if the patient was not tested and COVID-19 was not specified on the death certificate. To take this into account, \cite{DR} present a comprehensive framework for simultaneously modelling under-reporting and delayed-reporting. Extended here to include a spatial dimension, this is achieved by replacing Equation (1) in Section 3 of the main text with:
\begin{eqnarray}
x_{t,s}\mid \lambda_{t,s}, \theta_s &\sim & \mbox{Negative-Binomial}(\lambda_{t,s},\theta_s);\\
y_{t,s}\mid x_{t,s},\pi_{t,s} &\sim & \mbox{Binomial}(\pi_{t,s},y_{t,s});\\
\log\left(\frac{\pi_{t,s}}{1-\pi_{t,s}}\right) &=& i(t,s),
\end{eqnarray}
for $y_{t,s}\leq x_{t,s}$ and where $i(t,s)$ is a general function which may include covariates or random effects, e.g. covariates representing access to COVID-19 tests. The likelihood for $y_{t,s}$ is non-identifiable between a high $\lambda_{t,s}$ and a low $\pi_{t,s}$, or vice-versa, so in the case where all available counts are assumed potentially under-reported (i.e. $x_{t,s}$ is always unobserved), identifiability can be achieved using prior information \citep{CUR}.

\end{appendices}

\end{document}